\documentclass[aps,pre,twocolumn,showpacs,superscriptaddress,10pt]{revtex4}
\usepackage{graphicx}
\usepackage{subfigure}
\usepackage{amsmath}
\usepackage{amsfonts}
\usepackage{amssymb}
\usepackage{amsthm}
\usepackage{times}
\usepackage[colorlinks,citecolor=blue,linkcolor=blue]{hyperref}
\usepackage{color}

\begin{document}
\title{Efficiency at maximum power of a quantum Carnot engine with temperature tunable baths}

\newcommand{\MIT}{\affiliation{Department of Chemistry, Massachusetts Institute of Technology, 77 Massachusetts Avenue, Cambridge, MA 02139, USA}}
\newcommand{\Smart}{\affiliation{Singapore-MIT Alliance for Research and Technology (SMART) center, 1 CREATE Way, Singapore 138602, Singapore}}

\author{Junjie Liu}
\Smart
\MIT
\author{Chang-Yu Hsieh}
\Smart
\MIT
\author{Jianshu Cao}
\MIT
\Smart

\begin{abstract}
We investigate the efficiency at maximum power (EMP) of irreversible quantum Carnot engines that perform finite-time cycles between two temperature tunable baths. The temperature form we adopt can be experimentally realized in squeezed baths in the high temperature limit, which makes our proposal of practical relevance. Focusing on low dissipation engines, we first generalize the pervious upper as well as lower bounds for the EMP to temperature tunable cases in which they are solely determined by a generalized Carnot limit. As an illustrative example, we then consider a minimal heat engine model with a two-level spin as the working medium. It mimics a low dissipation engine as confirmed by finite time thermodynamic optimization results. The so-obtained EMP, being constrained by the generalized bounds, is well described by a generalized Curzon-Ahlborn efficiency as consequences of a left/right symmetry for a rate constant and low dissipations. Intriguing features of this minimal heat engine under optimal power output are also demonstrated.
\end{abstract}


\maketitle

\section{Introduction}
Heat engines, that convert heat into useful work \cite{Benenti.17.PR}, have been among central topics of thermodynamics since the seminal work of Carnot \cite{Carnot.18.NULL}. It is well-known that a quasi-static engine operating with two heat baths achieves the Carnot efficiency $\eta_C$. However, the power output (work per unit time) of such reversible engines vanish due to an infinitely long cycle time. Recent studies \cite{Allahverdyan.13.PRL,Shiraishi.16.PRL,Raz.16.PRL,Proesmans.16.PRL} suggest a general no-go theorem which rules out the possibility of heat engines with nonvanishing power and the Carnot efficiency simultaneously.

In recent years, considerable efforts have been devoted to irreversible heat engines operating with finite time cycles which produce finite power output at the expense of a reduced efficiency. Within the framework of finite-time thermodynamics \cite{Andresen.84.PT,Andresen.11.ACIE}, we can investigate the performance of finite-time, irreversible thermodynamic processes, and address the relation between efficiency and power. In particular, the efficiency at maximum power (EMP) of heat engines has attracted much attention. Theoretical developments based on linear as well as nonlinear irreversible thermodynamics \cite{Broeck.05.PRL,Esposito.09.PRL,Wang.12.EPL,Izumida.12.EPL,Proesmans.16.PRX,Proesmans.16.PRL}, the assumption of endoreversibility \cite{Curzon.75.AJP,Chen.89.JCP,Geva.92.JCP,Correa.14.PRE}, the low dissipation condition \cite{Schmiedl.08.EPL,Esposito.10.PRLa,Wang.12.PREa,Wang.12.PRE,Tomas.12.PRE,Broeck.13.EPL,Guo.13.PRE,Holubec.15.PRE,Gonzalez.17.PRE}, finite-sized heat baths \cite{Ondrechen.81.AJP,Izumida.14.PRL,Johal.16.EPL} , and molecular machines \cite{Seifert.11.PRL,Van.12.PRL,Golubeva.12.PRL} have been put forward. Among those proposals, the low dissipation Carnot engine has become paradigmatic due to the possibility of finding bounds for the EMP without any information on peculiarities of the heat transfer processes. In this model, irreversible contributions are assumed to be inversely proportional to the duration of the time spent on the heat transfer step, and particularities of the dissipations are englobed by dissipative coefficients, which are intrinsic properties of the engine \cite{Esposito.10.PRLa}. The Curzon-Ahlborn efficiency \cite{Curzon.75.AJP} is just a special limit of the resulting EMP and its bounds can apply to a wide set of experimental reported data. Furthermore, equivalences of low dissipation engines to nonlinear irreversible and endoreversible engines have been pointed out \cite{Izumida.12.EPL,Johal.17.PRE}.

Boosted by the advances in nano-fabrication, an intense theoretical effort has also been devoted to investigate quantum heat engines that exploit non-thermal baths \cite{Scully.03.S,Dillenschneider.09.EPL,Huang.12.PRE,Abah.14.EPL,Rossnagel.14.PRL,Perarnau.15.PRX,Alicki.15.NJP,Niedenzu.16.NJP,Manzano.16.PRE,Niedenzu.17.A,Klaers.17.A,Bijay.17.A}.  Non-thermal baths are quite common in nature, for instance the sunlight, continuous laser radiation \cite{Scully.97.NULL}, biological cells. Quantum reservoir engineering techniques \cite{Poyatos.96.PRL,Abah.12.PRL,Klaers.17.A} even enable the realization of non-thermal baths such as the squeezed thermal state \cite{Huang.12.PRE,Abah.14.EPL,Rossnagel.14.PRL,Alicki.15.NJP,Niedenzu.16.NJP,Manzano.16.PRE,Niedenzu.17.A,Bijay.17.A}. These stationary nonthermal baths are characterized by a temperature as well as additional parameters that quantify the degree of quantum coherence, quantum correlations, squeezing, etc. Surprisingly, the maximum efficiency of such nonequilibrium settings is limited by a generalized Carnot efficiency that can surpass the standard Carnot value $\eta_C$ \cite{Abah.14.EPL,Rossnagel.14.PRL}. Extracting work from a single non-thermal bath is also possible \cite{Scully.03.S,Abah.15.EPL,Klaers.17.A}. However, majority of theoretical studies on non-thermal engines are limited to the Otto cycle and its maximum efficiency, characteristics of non-thermal engines that perform Carnot cycles have been largely unexplored.

Inspired by the advances in nano-technologies, we focus on a particular kind of heat engines with non-thermal baths whose temperatures can be tunable \cite{Scully.03.S,Niedenzu.16.NJP,Niedenzu.17.A}. We choose the tunable temperature form as $T(1+2\sinh^2r)$ with $r$ the additional tuning parameter and $T$ the temperature of the bath, such an effective temperature form can be experimentally realized in squeezed thermal baths in the high temperature limit \cite{Klaers.17.A} with $r$ being the squeezing parameter. Hence our setup is of practical relevance. We consider a finite time quantum Carnot cycle where two \textquotedblleft isothermal" steps last for finite time intervals (the quotation marks on isothermal merely indicates that the working medium is in contact with a heat bath at constant effective temperature), and two quantum adiabatic steps whose time durations are negligible in comparison to the other time scales \cite{Schmiedl.08.EPL,Esposito.10.PRLa}.  

Working with low dissipation engines, we are able to derive general bounds for the EMP which are solely determined by a generalized Carnot limit. Previous results for thermal baths are recovered as a special case where tuning parameters of the hot and cold bath are equal.  We find that even the EMP can surpass the standard Carnot limit $\eta_C$ in certain parameter regimes. We further propose a minimal model of such quantum engines which can be implemented and tested experimentally using solid-state quantum devices. The working medium is a two-level spin which contacts the hot and cold baths with tunable temperatures alternately. By performing finite time thermodynamic optimization, we find this minimal engine indeed behaves as a low dissipation machine. The resulting EMP, being constrained by the general bounds, can be well described by a generalized Curzon-Ahlborn efficiency as consequences of a left/right symmetry for a rate constant and low dissipations. We also demonstrate that the optimal time durations in two "isothermal" processes equal, implying the dissipative coefficients behave inversely proportional to the bath tunable temperatures.

The paper is organized as follows. In section \ref{sec:1}, we study the EMP of general irreversible quantum Carnot engines with temperature tunable baths under the low dissipation assumption. In section \ref{sec:2}, we consider a minimal model and perform finite time thermodynamic optimization to obtain self-contained results. In section \ref{sec:3}, we summarize our findings and make final remarks.

\section{Low dissipation model: Efficiency at maximum power}\label{sec:1}
Consider a quantum heat engine in which the working medium contacts two heat baths alternately through a finite time quantum Carnot cycle, the tunable temperatures for the hot (H) and cold (C) bath read 
\begin{equation}\label{eq:tt}
T_H^e=T_H(1+2\sinh^2r_H),~~~T_C^e=T_C(1+2\sinh^2r_C), 
\end{equation}
respectively, with $T_H,T_C$ the thermodynamic temperatures and $r_H,r_C$ the tuning parameters of the corresponding bath. We require that $T_H>T_C$ and $r_H\ge r_C$. During a cycle, values of tuning parameters $r_H$ and $r_C$ are fixed such that "isothermal" steps can be defined. We also let $\hbar=1$ and $k_B=1$.

Let $t_C$ ($t_H$) be the time durations (finite but still sufficiently large compared with the relaxation time of the working medium) during which the working medium is in contact with the cold (hot) bath along a cycle. Noting an adiabatic transformation must be slow on the scale of the relaxation rate of the working medium. However, in principle, this relaxation can be arbitrarily fast, and thus the adiabatic transformation can also be made arbitrarily fast. Therefore, we make the assumptions that the time spent in the adiabatic steps of the irreversible quantum Carnot cycle is negligible compared to the times of the isothermal steps \cite{Schmiedl.08.EPL,Esposito.10.PRLa}.  Then the time duration of a cycle is just $t_H+t_C$. 

Working with the so-called low dissipation model, we actually consider a small departure from the reversible Carnot cycle by allowing weak dissipations through the finite time durations of the working medium with the baths. Therefore the absorbed heats during two finite time \textquotedblleft isothermal" steps can be expressed as \cite{Schmiedl.08.EPL,Esposito.10.PRLa,Cavina.17.A}
\begin{equation}\label{eq:pe}
Q_{H}=Q_0^H+\frac{Q_1^H}{t_H},~~~Q_{C}=Q_0^C+\frac{Q_1^C}{t_C},
\end{equation}
where $Q_0^H=T_H^e\Delta S$ and $Q_0^C=-T_C^e\Delta S$ are the contributions from quasi-static steps with $\Delta S$ the entropy change of the working medium, $Q_1^H<0$ and $Q_1^C<0$ are first order irreversible corrections due to finite durations. Then the power output reads
\begin{equation}
P~=~\frac{(T_H^e-T_c^e)\Delta S+Q_1^H/t_H+Q_1^C/t_C}{t_H+t_C}.
\end{equation}
In the quasi-static limit $t_C,t_H\to\infty$, the power output tends to zero and efficiency $\eta\equiv(Q_H+Q_C)/Q_H$ reaches the generalized Carnot limit \cite{Huang.12.PRE,Abah.14.EPL,Alicki.15.NJP}
\begin{equation}\label{eq:gc}
\eta_s=1-\frac{T_C^e}{T_H^e}.
\end{equation}
For equal tuning parameters $r_H=r_C$, we recover the standard Carnot limit $\eta_C=1-T_C/T_H$.

For low dissipation engines with maximum power output, we can find bounds for the EMP 
\begin{equation}\label{eq:bound}
\frac{\eta_s}{2}\equiv\eta_{min}^{\ast}\le\eta^{\ast}\le\eta_{max}^{\ast}\equiv\frac{\eta_s}{2-\eta_s}.
\end{equation}
This result can be regarded as a generalization of bounds obtained for low dissipation engines with thermal baths [Ref. \cite{Esposito.10.PRLa}] to that with temperature tunable baths. For equal tuning parameters $r_H=r_C$, $\eta_s$ becomes the standard Carnot limit $\eta_C$, we then recover previous results for thermal baths.

\begin{figure}[tbh!]
  \centering
  \includegraphics[width=1\columnwidth]{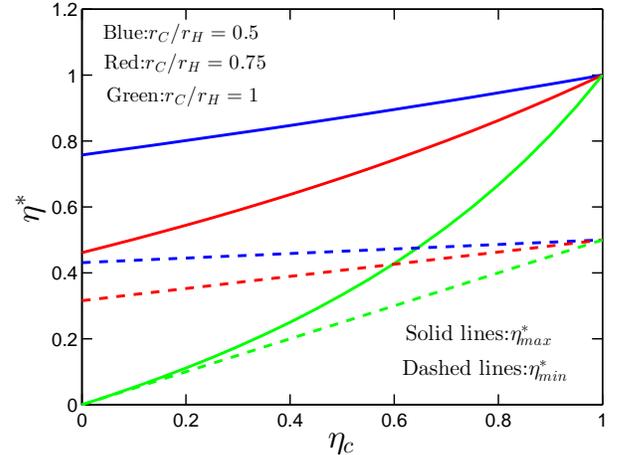}
\caption{(Color online) Bounds for the EMP as a function of standard Carnot limit $\eta_C$ with different tuning parameters. Solid lines represent $\eta^{\ast}_{max}$, dashed lines denote $\eta^{\ast}_{min}$. Different ratios $r_C/r_H$ are marked with different colors (from top to bottom with one type of lines: $r_C/r_H=0.5$, $r_C/r_H=0.75$, $r_C/r_H=1$). We fix $r_H=2$.}
\label{fig:bound}
\end{figure}

To examine the effect of tuning parameters, we plot bounds with different ratios $r_C/r_H$ in Fig. \ref{fig:bound}. As can be seen, for equal tuning parameters with $r_C/r_H=1$, the bounds reduce to previous results \cite{Esposito.10.PRLa} and the EMP can never exceed the standard Carnot limit $\eta_C$. However, for smaller ratios $r_C/r_H$, there are regions where the EMP can surpass the standard Carnot limit $\eta_C$. 

\section{A minimal model}\label{sec:2}
\subsection{Setup}
To illustrate the above general results, we consider a minimal heat engine model which consists of a single two-level spin weakly interacting with two temperature tunable heat baths alternatively. The temperature tunable heat bath can be realized by squeezed baths in the high temperature limit \cite{Klaers.17.A}. The engine is carried through the cycle by an external driving protocol in the working medium's Hamiltonian
\begin{equation}
H_s(t)~=~\frac{\Delta(t)}{2}\sigma_z.
\end{equation}
The state of the working medium is specified by its population $p_z(t)\equiv\langle\sigma_z\rangle$. By adopting the wide band approximation, the equation of motion of $p_z(t)$ for the working medium in contact with a heat bath at temperature $T^e=T(1+2\sinh^2r)$ satisfies the following Redfield master equation \cite{Hofer.17.A,Gonzalez.17.A}
\begin{equation}\label{eq:eom}
\dot{p}_z(t)~=~-\Gamma\coth\frac{\Delta(t)}{2T^e}p_z(t)-\Gamma,
\end{equation}
where $\Gamma$ is a rate constant, the dot is a symbolic notation for the partial time derivative. Without loss of generality, in the following discussion, we limit ourselves to a left/right symmetry $\Gamma_H=\Gamma_C=\Gamma$, namely, the exchange rates of the working medium with hot and cold baths are equal. And the high temperature limit with $T\gg\Delta$ is considered.

For weak couplings, the quantum thermodynamics of the working medium is well-defined. The internal energy of the working medium reads
\begin{equation}
E(t)~=~\frac{\Delta(t)}{2}p_z(t)
\end{equation} 
Then the first law of thermodynamics can be expressed as
\begin{equation}
\dot{E}(t)~=~\dot{W}(t)+\dot{Q}(t),
\end{equation}
where we make the following identification for rates of work $W$ and heat $Q$
\begin{equation}\label{eq:wq}
\dot{W}(t)=\frac{\dot{\Delta}(t)}{2}p_z(t),~~~\dot{Q}(t)=\frac{\Delta(t)}{2}\dot{p}_z(t).
\end{equation}
We adopt the conventions that positive work is done on the working medium and that positive heat indicates the working medium absorbs heat from the heat bath. It is worthwhile to mention that Eq. (\ref{eq:wq}) is applicable to both the thermal equilibrium cases as well as non-equilibrium cases.

\begin{figure}[tbh!]
  \centering
  \includegraphics[width=0.85\columnwidth]{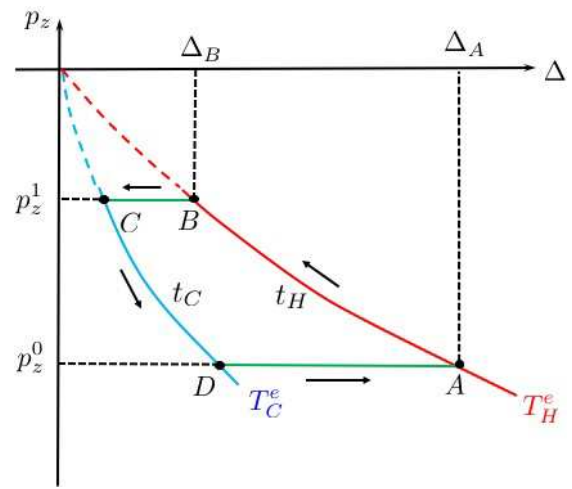} 
\caption{(Color online) A finite time quantum Carnot cycle consisting of two finite time "isothermal" steps ($A\to B, C\to D$) with time durations $t_H, t_C$ and two quantum adiabatic steps ($B\to C, D\to A$). $p_z^0$ and $p_z^1$ are spin populations, $\Delta_A$ and $\Delta_B$ are energy gaps of the working medium at state $A$ and $B$, respectively. The tunable temperatures of hot and cold baths are denoted as $T_H^e$ and $T_C^e$, respectively. Positive work output is obtained by going anti-clockwise.}
\label{fig:FC}
\end{figure}

The operating cycle is a finite time Carnot cycle as depict in Fig. \ref{fig:FC}.  There $A\to B$ and $C\to D$ are two finite time "isothermal" steps with time durations $t_H$ and $t_C$ during which the working medium is in contact with heat baths with tunable temperature $T_H^e$ and $T_C^e$, respectively.  We require that the working medium are in stationary states with heat baths at states $i$ with $i = A,B,C,D$.
To construct quantum adiabatic steps ($B\to C, D\to A$), we should have the following scaling relations \cite{Quan.07.PRE,Xiao.15.PRE}
\begin{equation}\label{eq:si}
\frac{\Delta_B}{\Delta_C}~=~\frac{T_H^e}{T_C^e},~~~\frac{\Delta_A}{\Delta_D}~=~\frac{T_H^e}{T_C^e}
\end{equation}
Therefore, we can ensure that the population of the working medium remains constant during the adiabatic steps with
\begin{equation}\label{eq:bc}
p_z^1=-\tanh\frac{\Delta_B}{2T_H^e},~~~p_z^0=-\tanh\frac{\Delta_A}{2T_H^e}.
\end{equation}

\subsection{Optimal operation}
For such a minimal heat engine, our aim is now to find an optimal driving protocol $\Delta(t)$ which maximizes the power output and yields the corresponding EMP. Within the framework of finite time thermodynamics \cite{Andresen.84.PT,Andresen.11.ACIE}, such a finite time thermodynamic optimization can be done in two steps \cite{Esposito.10.EPL,Esposito.10.PRE}: First, we maximize the absorbed heat in two finite time "isothermal" steps with fixed time durations and boundary conditions in the functional space of $\Delta(t)$. By doing so, we will find an optimal driving protocol $\Delta(t)$. Next, we further maximize the power output with respect to time durations. With the optimal driving protocol and optimal time durations, we can obtain the EMP for this minimal heat engine. 

The optimization problem involved in the first step is nontrivial as optimal protocols exhibit sudden jumps in general \cite{Esposito.10.EPL,Esposito.10.PRE,Schmiedl.07.PRL,Schmiedl.08.EPL,Then.08.PRE,Alex.08.JCP,Izumida.08.EPL}. To overcome this difficulty, we express the time-dependent energy gap $\Delta(t)$ as a functional of the population $p_z(t)$ which by definition is always differentiable according to the equation of motion Eq. (\ref{eq:eom})
\begin{equation}
\Delta[p_z]~=~T^e\ln\frac{\Gamma+\dot{p}_z-\Gamma p_z}{\Gamma+\dot{p}_z+\Gamma p_z},
\end{equation}
where $T^e$ can be either $T_H^e$ or $T_C^e$. Inserting the above functional into the definition of heat [Eq. (\ref{eq:wq})], we find
\begin{equation}
Q[p_z]~=~\int_0^{\tau}\mathcal{L}(p_z,\dot{p}_z)dt,
\end{equation}
where $\tau$ can be either $t_H$ or $t_C$ with $Q$ corresponds to $Q_H$ and $Q_C$, respectively, and $\mathcal{L}(p_z,\dot{p}_z)=\frac{1}{2}T^e\dot{p}_z\ln\frac{\Gamma+\dot{p}_z-\Gamma p_z}{\Gamma+\dot{p}_z+\Gamma p_z}$. Therefore, the optimization problem turns into one which needs to find an optimal evolution $p_z(t)$ under the constraints of given initial and final stationary populations, from which we then identify the corresponding optimal driving protocol $\Delta(t)$.

The maximum value of absorbed heat $Q$ can be obtained via the Euler-Lagrange equation which yields
\begin{equation}
\mathcal{L}-\dot{p}_z\frac{\partial\mathcal{L}}{\partial \dot{p}_z}=\tilde{K}
\end{equation}
with $\tilde{K}$ a constant of integration. Introducing $\dot{\tilde{p}}_z=\dot{p}_z/\Gamma$ and $K=-\tilde{K}/(\Gamma T^e)$, we find
\begin{equation}\label{eq:ELE}
\frac{\dot{\tilde{p}}_z^2p_z}{(1+\dot{\tilde{p}}_z)^2-p_z^2}=K.
\end{equation}
Obviously, from the above equations, we know that $K$ is determined by boundary conditions as well as model's parameters. 

Before turning to the solution of Eq. (\ref{eq:ELE}), we first examine the physical consequence of the constant $K$. To see that, we insert Eq. (\ref{eq:eom}) into Eq. (\ref{eq:ELE}) and obtain a quadratic equation for $p_z$
\begin{equation}
\left[p_z\coth\frac{\Delta(t)}{2T^e}+1\right]^2=Kp_z\mathrm{csch}^2\frac{\Delta(t)}{2T^e},
\end{equation}
from which we deduce $K<0$ in general and $K=0$ only for quasi-static isothermal steps. Hence $K$ measures how far the state of the system deviates from the quasi-static limit \cite{Esposito.10.EPL}. To ensure that the low dissipation condition is fulfilled by our minimal heat engine,  values of $K$ should be small (see Fig. \ref{fig:tt} (a)). Solving the above quadratic equation, we find
\begin{eqnarray}
p_{z,\pm}(t)&=&-\tanh\frac{\Delta(t)}{2T^e}+K\mathrm{sech}^2\frac{\Delta(t)}{2T^e}\nonumber\\
&&\times\left[1\pm\sqrt{1-\frac{4}{K}\sinh\frac{\Delta(t)}{2T^e}\cosh\frac{\Delta(t)}{2T^e}}\right].
\end{eqnarray}
For a finite time driving process, we easily find $\lim_{t\to0}p_z(t)\neq p_z(0)=-\tanh\frac{\Delta(0)}{2T^e}$. This apparent inconsistency indicates that $\Delta(0^{+})\neq\Delta(0)$, namely, there must be a sudden jump from $\Delta(0)$ to $\Delta(0^+)$ (similar situation happens for $\Delta(\tau^-)$ and $\Delta(\tau)$), in consistent with pervious findings \cite{Esposito.10.EPL,Esposito.10.PRE,Schmiedl.07.PRL,Schmiedl.08.EPL,Then.08.PRE,Alex.08.JCP,Izumida.08.EPL}. Besides the sudden jumps at the beginning and end of the "isothermal" processes, we should mention that during the processes the change of $\Delta$ is rather smooth and slow. Furthermore, the solution with minus sign indicates the population increases from its initial value as time increases and thus should be used for the finite time "isothermal" step in contact with the hot bath $T^e=T_H^e$ (see Fig. \ref{fig:FC}). Similarly, the solution with plus sign should be used for the finite time "isothermal" step with the cold bath $T^e=T_C^e$ in which the population decreases as time increases.

Next we return to the discussion of Eq. (\ref{eq:ELE}) and directly solve the quadratic equation of $\dot{\tilde{p}}_z$ by noting $K<0$, which yields two possibilities
\begin{equation}\label{eq:pdot}
\dot{\tilde{p}}_{z,\pm}(t)~=~\frac{-K\pm\sqrt{Kp_z\left[Kp_z+1-p_z^2\right]}}{K-p_z}.
\end{equation}
Similarly, we can identify that $\dot{\tilde{p}}_{z,-}(t)<0$ describes the finite time "isothermal" step with the cold bath, $\dot{\tilde{p}}_{z,+}(t)>0$ represents the finite time "isothermal" step with the hot bath. Using those facts together with the boundary conditions Eq. (\ref{eq:bc}), we can obtain equations that determine constant $K$:
\begin{eqnarray}
\Gamma t_H &=& \left.\int_{p_z^0}^{p_z^1}\frac{dp_z}{\dot{\tilde{p}}_{z,+}}\right|_{K=K_H},\label{eq:kh}\\
\Gamma t_C &=& \left.\int_{p_z^1}^{p_z^0}\frac{dp_z}{\dot{\tilde{p}}_{z,-}}\right|_{K=K_C},\label{eq:kc}
\end{eqnarray}
where $p_z^0$ and $p_z^1$ are given by Eq. (\ref{eq:bc}). Since the above integrals have no analytical expressions, we should solve them numerically. By doing so, we can obtain values of $K_H$ and $K_C$ from Eqs. (\ref{eq:kh}) and (\ref{eq:kc}) for the "isothermal" steps with hot and cold baths, respectively. 

With $K_H$, $K_C$ and Eq. (\ref{eq:pdot}), we can evaluate absorbed heat in two optimal finite time "isothermal" steps with fixed time durations $t_H$ and $t_C$
\begin{eqnarray}
Q_H &=& \frac{1}{2}\int_0^{t_H}\Delta(t)\dot{p}_z(t)dt~=~\frac{1}{2}\int_{p_z^0}^{p_z^1}\Delta[p_z]dp_z\nonumber\\
&=& \left.\frac{T_H^e}{2}\int_{p_z^0}^{p_z^1}\ln\frac{1+\dot{\tilde{p}}_{z,+}-p_z}{1+\dot{\tilde{p}}_{z,+}+p_z}dp_z\right|_{K=K_H},\label{eq:qh}\\
Q_C &=& \left.\frac{T_C^e}{2}\int_{p_z^1}^{p_z^0}\ln\frac{1+\dot{\tilde{p}}_{z,-}-p_z}{1+\dot{\tilde{p}}_{z,-}+p_z}dp_z\right|_{K=K_C}.\label{eq:qc}
\end{eqnarray}
Then we vary the values of $t_H$ and $t_C$ and numerically solve Eqs. (\ref{eq:kh})-(\ref{eq:qc}) repeatedly to get a maximum value of power output $P=(Q_H+Q_C)/(t_H+t_C)$. 

\subsection{Numerical results}
\begin{figure}[tbh!]
  \centering
  \includegraphics[width=1\columnwidth]{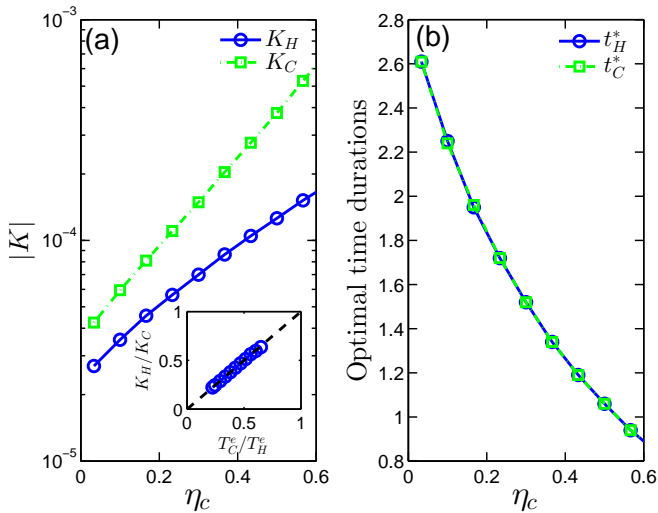}
\caption{(Color online) (a) Absolute values of constants $K_H$ (blue circles) and $K_C$ (green squares), (b) Optimal time durations $t_H^{\ast}$ (blue circles) and $t_C^{\ast}$ (green squares). We choose $r_H=2$, $r_C=1.8$, $\Delta_A=5$meV, $\Delta_B=3$meV, $T_H=25.8$meV, $T_C$ varies from 9.46meV to 24.94meV, $\Gamma=0.005$.}
\label{fig:tt}
\end{figure}

A set of numerical results of optimal time durations $t_H^{\ast}$, $t_C^{\ast}$ as well as constants $K_H$ and $K_C$ which leads to maximum power output is shown in Fig. \ref{fig:tt}. 
From the figure, several intriguing features of our minimal model should be remarked: (a) Constants $K_H$ and $K_C$ which measure the deviations from the quasi-static limit are quite small, implying that the minimal engine studied here mimics a low dissipation model and hence the first order perturbation expansions Eq. (\ref{eq:pe}) is well justified. Their ratio satisfies a simple relation $K_H/K_C=T_C^e/T_H^e$ (see the inset of Fig. \ref{fig:tt}(a)) since the cold isothermal optimal protocol is the time reversal of the hot one when considering maximum power output \cite{Cavina.17.A}. (b) Optimal time durations $t_H^{\ast}$ and $t_C^{\ast}$ are equal due to time-reversal driving protocols and the left/right symmetry of the rate constant $\Gamma$. 

\begin{figure}[tbh!]
  \centering
  \includegraphics[width=1\columnwidth]{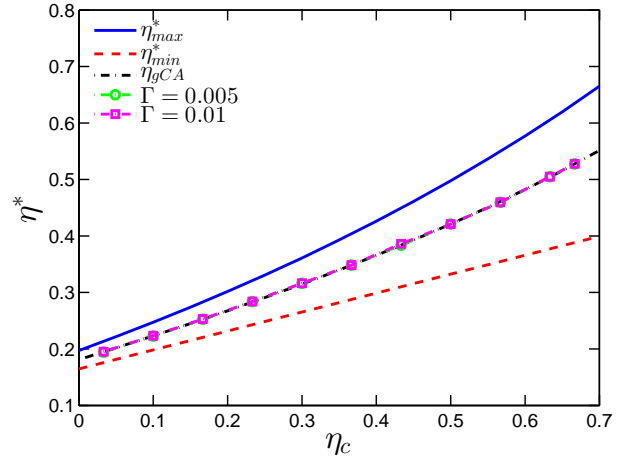}
\caption{(Color online) EMP $\eta^{\ast}$ as a function of standard Carnot limit $\eta_C$. Blue solid line is the upper bound $\eta_{max}^{\ast}$ (see Eq. (\ref{eq:bound})), red dashed line is the lower bound $\eta_{min}^{\ast}$ (see Eq. (\ref{eq:bound})), black dashed-dotted line is the generalized Curzon-Ahlborn efficiency (see Eq. (\ref{eq:gca})), green circles and maroon squares are EMP results of the minimal model with $\Gamma=0.005$ and $\Gamma=0.01$, respectively. We choose $r_H=2$, $r_C=1.8$, $\Delta_A=5$meV, $\Delta_B=3$meV, $T_H=25.8$meV, $T_C$ varies from 8.6meV to 24.94meV.}
\label{fig:eta}
\end{figure}

With the optimal parameters and driving protocols, we can evaluate the EMP as
\begin{equation}
\eta^{\ast}~=~\frac{Q_H+Q_C}{Q_H},
\end{equation}
Results of the EMP for this minimal model are presented in Fig. \ref{fig:eta}. From the figure, it is evident that the EMP is constrained by the general bounds in Eq. (\ref{eq:bound}). We also note that the EMP is independent of values of the exchange rate constant $\Gamma$. Actually, on varying $\Gamma$, the optimal time durations $t_H^{\ast}$ and $t_C^{\ast}$ change in such a way that right-hand-sides of Eqs. (\ref{eq:kh}) and (\ref{eq:kc}) remain unchanged, resulting in the same $K_H$ and $K_C$. According to  Eqs. (\ref{eq:qh}) and (\ref{eq:qc}), we will obtain the same maximum values of absorbed heat and consequently the same EMP. More Interestingly, the EMP of this minimal heat engine can be well described by a generalized Curzon-Ahlborn (gCA) efficiency
\begin{equation}\label{eq:gca}
\eta_{gCA}~=~1-\sqrt{1-\eta_s}
\end{equation}
with $\eta_s$ the generalized Carnot limit Eq. (\ref{eq:gc}). 
This agreement results from the left/right symmetry of the rate constant $\Gamma$ and the low dissipation regime we considered, as noted in Refs. \cite{Esposito.10.PRE}. Thus, for genuine heat engines with a controllable Carnot limit, the EMP still has the universality.

\section{Conclusions}\label{sec:3}
In this study, we focus on the efficiency at maximum power (EMP) of heat engines fuelled by non thermal baths whose temperatures are adjustable. The tunable temperature form we chose can be experimentally realized in squeezed baths in the high temperature limit, which makes our investigation of practical relevance. 

Considering low dissipation machines, we are able to derive general upper and lower bounds for the EMP which are solely determined by a generalized Carnot limit. Those bounds reduce to previous results provided that the tuning parameters in hot and cold baths are equal. With unequal tuning parameters, we find that even the EMP can surpass the standard Carnot limit in certain parameter regimes. To illustrate general results, we consider a minimal heat engine model which utilizes a two-level spin as the working medium, it mimics a low dissipation engine as confirmed by finite time thermodynamic optimization results. The resulting EMP, being constrained by the general bounds, is well described by a generalized Curzon-Ahlborn efficiency as consequences of a left/right symmetry for a rate constant and low dissipations. 
In future works, we wish to address effects of coupling strength on the EMP of this minimal model \cite{Xu.16.NJP,Newman.17.PRE}.

\begin{acknowledgments}
J. Liu thanks Prof. Dario Poletti and Chen Wang for valuable discussions. J. Liu and C. Hsieh acknowledge the support from the Singapore-MIT Alliance for Research and Technology (SMART), J. Cao is supported by NSF (grant no. CHE-1112825) and SMART.
\end{acknowledgments}


\end{document}